\documentclass[aps,twocolumn,showpacs,preprintnumbers,floats]{revtex4}\def\@cite#1#2{\textsuperscript{[{#1\if@tempswa , #2\fi}]}}

\usepackage{graphicx}
\usepackage{amsmath}
\usepackage{amsfonts}
\usepackage{amssymb}
\usepackage{color}
\usepackage{subfigure}
\usepackage{epsfig}
\usepackage{morefloats}
\usepackage{multirow}
\usepackage{graphicx,booktabs}
\usepackage{mathrsfs}
\usepackage{txfonts}
\usepackage{indentfirst}
\usepackage{graphicx,booktabs}

\usepackage{longtable,lscape}

\newcommand{\vsig}{\mbox{\boldmath$\sigma$\unboldmath}}

\newcommand{\vell}{\mbox{\boldmath$\ell$\unboldmath}}

\usepackage{comment}

\begin{document}

\title{Interpretation of the newly observed $\Sigma_b(6097)^{\pm}$ and $\Xi_b(6227)^-$ states as the $P$-wave bottom baryons}
\author{
Kai-Lei Wang$^{1}$, Qi-Fang L\"{u}$^{1,2}$~\footnote {E-mail: lvqifang@hunnu.edu.cn}, Xian-Hui Zhong$^{1,2}$~\footnote {E-mail: zhongxh@hunnu.edu.cn}}

\affiliation{ 1) Department
of Physics, Hunan Normal University, and Key Laboratory of
Low-Dimensional Quantum Structures and Quantum Control of Ministry
of Education, Changsha 410081, China }

\affiliation{ 2) Synergetic Innovation Center for Quantum Effects and Applications (SICQEA),
Hunan Normal University, Changsha 410081, China}

\begin{abstract}

The strong decays of the $P$-wave singly bottom heavy baryons belonging to $\mathbf{6}_F$ are investigated with a constituent quark model in the $j$-$j$ coupling scheme.  The results show that the newly observed $\Sigma_b(6097)$ and $\Xi_b(6227)$ states by the LHCb collaboration can be assigned as the $\lambda$-mode $P$-wave bottom baryons belonging to $\mathbf{6}_F$. Given the heavy quark symmetry, both the $\Sigma_b(6097)$ and $\Xi_b(6227)$ states favor the light spin $j=2$ states with spin-parity numbers $J^P=3/2^-$ or $J^P=5/2^-$. More $P$-wave baryons in the $\Sigma_b$, $\Xi_b'$, and $\Omega_b$ families are most likely to be observed in future experiments for their relatively narrow width.
\end{abstract}

\maketitle

\section{Introduction}{\label{introduction}}

Recently, the LHCb collaboration announced the observation of a
new bottom baryon $\Xi_b(6227)^-$ in both
$\Lambda_b^0K^-$ and $\Xi_b^0\pi$ decay modes~\cite{Aaij:2018yqz}. Just several weeks ago, the LHCb collaboration found two new resonances $\Sigma_b(6097)^{\pm}$
in the $\Lambda_b^0 \pi^{\pm}$ channels~\cite{Aaij:2018tnn}. The measured masses and widths of the $\Xi_b(6227)^-$ and $\Sigma_b(6097)^{\pm}$ are presented as follows,
\begin{eqnarray}
m[\Xi_b(6227)^-] = 6226.9\pm2.0\pm0.3\pm0.2~\rm{MeV},
\end{eqnarray}
\begin{eqnarray}
\Gamma[\Xi_b(6227)^-] = 18.1\pm5.4\pm1.8~\rm{MeV},
\end{eqnarray}
\begin{eqnarray}
m[\Sigma_b(6097)^-] = 6098.0\pm1.7\pm0.5~\rm{MeV},
\end{eqnarray}
\begin{eqnarray}
\Gamma[\Sigma_b(6097)^-] = 28.9\pm4.2\pm0.9~\rm{MeV},
\end{eqnarray}
\begin{eqnarray}
m[\Sigma_b(6097)^+] = 6095.8\pm1.7\pm0.4~\rm{MeV},
\end{eqnarray}
\begin{eqnarray}
\Gamma[\Sigma_b(6097)^+] = 31.0\pm5.5\pm0.7~\rm{MeV}.
\end{eqnarray}
Compared with the mass spectrum predicted in various models~\cite{Capstick:2000qj,Ebert:2011kk,Ebert:2007nw,Ebert:2005xj,Yoshida:2015tia,Roberts:2007ni,Valcarce:2008dr,
Karliner:2008sv,Chen:2014nyo,Mao:2015gya,Wang:2010it}, it is found that the $\Xi_b(6227)^-$ and $\Sigma_b(6097)^{\pm}$ are good candidates of the $\lambda$-mode $P$-wave bottom baryons belonging to $\mathbf{6}_F$.

To explain the nature of $\Xi_b(6227)^-$ and $\Sigma_b(6097)^{\pm}$, recently the strong decays of the excited singly bottom states were studied with the $^3P_0$ model in Ref.~\cite{Chen:2018orb,Chen:2018vuc}. The results indicate that $\Xi_b(6227)^-$ and $\Sigma_b(6097)^{\pm}$ may be good candidates of the $P$-wave states with $J^P=3/2^-$ or $J^P=5/2^-$ in the $j$-$j$ coupling scheme. In Ref.~\cite{Wang:2017kfr}, we studied the strong decays of the singly bottom baryons within the chiral quark model in the $L$-$S$ coupling scheme. Comparing the predictions with the observations, one can find that both $\Xi_b(6227)$ and $\Sigma_b(6097)$ could be assigned as the $P$-wave states with $J^P=3/2^-$ or $J^P=5/2^-$ as well. However, the partial width ratios and decay modes of some $P$-wave $\Xi_b'$ states predicted within the $j$-$j$ coupling scheme are different from those predicted within the $L$-$S$ coupling scheme.

We have noted that for the singly heavy baryons there may exist some effects of heavy quark symmetry, which was not carefully considered in our previous work~\cite{Wang:2017kfr}. To approximately preserve the heavy quark symmetry with the finite heavy quark masses, the physical states may be closer to the $j$-$j$ coupling basis. In this coupling scheme, the states are mixed states between the states of the $L$-$S$ coupling scheme with the same spin-parity ($J^P$) numbers. It is interesting to find that the newly observed state $\Omega_c(3000)$ at LHCb~\cite{Aaij:2017nav} may be explained as a $J^P=1/2^-$ resonance via a $|^2P_{1/2}\rangle$-$|^4P_{1/2}\rangle$ mixing~\cite{Wang:2017hej}. This may be evidence that the physical states of singly heavy baryons more favor the $j$-$j$ coupling scheme. In another word, the configuration mixing effects between the singly heavy baryon states with the same $J^P$ should be considered if one adopts the $L$-$S$ coupling scheme.

In this work, considering the requirements of the heavy quark symmetry we study the strong decay behaviors of the $\lambda$-mode $P$-wave bottom baryons $\Sigma_b$, $\Xi_b'$ and $\Omega_b$ in the $j$-$j$ coupling scheme with the chiral quark model. In this coupling scheme, the states with light spin $j=0$ and $j=1$ cannot decay into some special final states, which shows selection rules. The decay modes together with the predicted total widths suggest that both the $\Xi_b(6227)^-$ and $\Sigma_b(6097)^{\pm}$ are good candidates of light spin $j=2$ states, while the $J^P=3/2^-$ and $J^P=5/2^-$ cannot be distinguished due to lack of data. The predicted narrow widths of other flavor symmetric partners can provide helpful information for future experiments.

This paper is organized as follows. The spectrum and notations are presented in Sec.~\ref{spectrum}. The chiral quark model is briefly introduced in Sec.~\ref{model}. The  strong decays of the $\lambda$-mode $P$-wave $\Sigma_b$, $\Xi_b'$ and $\Omega_b$ are estimated in Sec.~\ref{results}. A short summary is presented in the last section.


\section{SPECTROSCOPY }\label{spectrum}

The heavy baryon containing a heavy quark violates
the SU(4) symmetry. However, the SU(3) symmetry between the other two
light quarks ($u$, $d$, or $s$) is approximately kept. According to
the symmetry, the heavy baryons containing a single heavy quark
belong to two different SU(3) flavor representations: the symmetric sextet $\mathbf{6}_F$ and
antisymmetric antitriplet $\bar{\mathbf{3}}_F$~\cite{Wang:2017kfr}.
For example, the $\Sigma_b^{-,0,+}$, $\Xi_b'^{0,-}$ and $\Omega_b$ form a $\mathbf{6}_F$ representation,
while $\Lambda_b$ and $\Xi_b^{0,-}$ form a $\bar{\mathbf{3}}_F$ representation.

In the $L$-$S$ coupling scheme, the quark model states are
constructed by~\cite{Roberts:2007ni}
\begin{equation}
\left|^{2S+1}L_{J}\right\rangle = \left|\left[\left(\ell_\rho \ell_\lambda\right)_L\left(s_{\rho}s_Q\right)_S\right]_{J^P}\right\rangle,
\end{equation}
where $\ell_\rho$ and $\ell_\lambda$ correspond to the eigenvalues of
the orbital angular momenta $\vell_\rho$ and $\vell_\lambda$ for the $\rho$ and $\lambda$ oscillators, respectively,
and $L$ stands for the eigenvalue of the total orbital angular momentum $\mathbf{L}=\vell_\rho+\vell_\lambda$~\cite{Zhong:2007gp,Wang:2017hej}. The parity of a state is determined by $P=(-1)^{\ell_\rho+\ell_\lambda}$. Quantum numbers $s_{\rho}$ and $s_Q$ correspond to the eigenvalues
of the total spin of the two light quarks $\mathbf{s}_{\rho}$ and the spin of the heavy quark $\mathbf{s}_Q$, respectively,
while $S$ stands for the eigenvalue of the total spin angular momentum $\mathbf{S}=\mathbf{s}_{\rho}+\mathbf{s}_Q$. According to the quark model classification, there are five $\lambda$-mode $1P$-wave states belonging
to $\mathbf{6}_F$~\cite{Wang:2017kfr}. These states and their corresponding quantum numbers have been collected in Table~\ref{LS}.

\begin{table}[htp]
\begin{center}
\caption{\label{LS} The classifications of the low-lying $P$-wave states belonging to $\mathbf{6}_F$ in the $L$-$S$ coupling scheme. The states in the $L$-$S$ coupling scheme are denoted by $|^{2S+1}L_{J} \rangle$. }
\begin{tabular}{p{1.8cm}p{0.6cm}p{0.6cm}p{0.6cm}p{0.6cm}p{0.6cm}p{0.6cm}p{0.6cm}p{0.6cm}p{0.6cm}}  \hline \hline
$~~~~~|^{2S+1}L_{J} \rangle$        &$J^P$              &$\ell_{\rho}$    &$\ell_{\lambda}$    &$L$   &$s_{\rho}$    &$s_{Q}$                    &~~$S$\\ \hline
$~~~~~|^2P_{1/2} \rangle$  &$\frac{1}{2}^-$           & 0                &1          &1       &1         &$\frac{1}{2}$       &~~$\frac{1}{2}$  \\
$~~~~~|^2P_{3/2} \rangle$  &$\frac{3}{2}^-$             & 0                &1          &1       &1         &$\frac{1}{2}$       &~~$\frac{1}{2}$      \\
$~~~~~|^4P_{1/2} \rangle$  &$\frac{1}{2}^-$            & 0                &1          &1       &1         &$\frac{1}{2}$       &~~$\frac{3}{2}$     \\
$~~~~~|^4P_{3/2} \rangle$  &$\frac{3}{2}^-$           & 0                &1          &1       &1         &$\frac{1}{2}$       & ~~$\frac{3}{2}$   \\
$~~~~~|^4P_{5/2} \rangle$  &$\frac{5}{2}^-$           & 0                &1          &1       &1         &$\frac{1}{2}$       & ~~$\frac{3}{2}$     \\
\hline\hline
\end{tabular}
\end{center}
\end{table}

\begin{table}[htp]
\begin{center}
\caption{\label{JJ}  The classifications of the low-lying $P$-wave states belonging to $\mathbf{6}_F$ in the $j$-$j$ coupling scheme. The states in the $j$-$j$ coupling scheme are denoted by $|J^P,j \rangle$.}
\begin{tabular}{p{2.5cm}p{0.6cm}p{0.6cm}p{0.6cm}p{0.6cm}p{0.6cm}p{0.6cm}p{0.6cm}p{0.6cm}p{0.6cm} |}\hline\hline
$|J^P,j \rangle$      &$J^P$            &$j$   &$\ell_{\rho}$    &$\ell_{\lambda}$    &$L$   &$s_{\rho}$    &$s_{Q}$                     \\ \hline
$|J^P=\frac{1}{2}^-,0\rangle$   &$\frac{1}{2}^-$  & 0        & 0                &1          &1       &1         &$\frac{1}{2}$         \\
$|J^P=\frac{1}{2}^-,1\rangle$   &$\frac{3}{2}^-$  & 1        & 0                &1          &1       &1         &$\frac{1}{2}$              \\
$|J^P=\frac{3}{2}^-,1\rangle$   &$\frac{1}{2}^-$  & 1        & 0                &1          &1       &1         &$\frac{1}{2}$           \\
$|J^P=\frac{3}{2}^-,2\rangle$   &$\frac{3}{2}^-$  & 2        & 0                &1          &1       &1         &$\frac{1}{2}$            \\
$|J^P=\frac{5}{2}^-,2\rangle$   &$\frac{5}{2}^-$  & 2        & 0                &1          &1       &1         &$\frac{1}{2}$            \\
\hline\hline
\end{tabular}
\end{center}
\end{table}

It should be pointed out that the heavy-quark symmetry as an approximation is commonly adopted for the study of the singly heavy baryons.
In the heavy quark effective theory description, the spin of the heavy quark $\mathbf{s}_Q$ and the total angular momentum of the two light quarks $\mathbf{j}=\mathbf{L}+\mathbf{s}_{\rho}$ are separately conserved~\cite{Cheng:2015iom}. The total angular momentum is given by $\mathbf{J}=\mathbf{j}+\mathbf{s}_Q$.
In the heavy-quark symmetry limit, the quark model states may more favor the $j$-$j$ coupling scheme
\begin{eqnarray}
\left|J^P,j\right\rangle = \left|\left\{\left[\left(\ell_\rho \ell_\lambda\right)_L s_{\rho}\right]_js_Q\right\}_{J^P}\right\rangle.
\end{eqnarray}
The five $1P$-wave states belonging to $\mathbf{6}_F$ in the $j$-$j$ coupling scheme and
their corresponding quantum numbers have been collected in Table~\ref{JJ}.

The states of the $j$-$j$ coupling scheme are linear combinations of the states of the
$L$-$S$ coupling scheme. The precise relationship is~\cite{Roberts:2007ni}
\begin{eqnarray}\label{Rela}
\left|\left\{\left[\left(\ell_\rho \ell_\lambda\right)_Ls_{\rho}\right]_js_Q\right\}_{J^P}\right\rangle =(-1)^{L+s_\rho+J+\frac{1}{2}}\sqrt{2j+1} \sum_{S}\sqrt{2S+1}  ~~\nonumber\\
\begin{Bmatrix}L &s_\rho &j\\s_Q &J &S \end{Bmatrix} \left|\left[\left(\ell_\rho \ell_\lambda\right)_L\left(s_{\rho}s_Q\right)_S\right]_{J^P}\right\rangle.~~~~~~~~~
\end{eqnarray}
The heavy-quark symmetry may tell us that there are configuration mixing between the singly heavy baryon states
with the same $J^P$ numbers in the $L$-$S$ coupling scheme. In the heavy quark limit, the mixing angles
are determined by Eq.(\ref{Rela}).


\section{ Strong decay  }\label{model}

To study the strong decays of the low-lying $P$-wave bottom baryons,
we apply the chiral quark model~\cite{Manohar:1983md}.
In this model, the light pseudoscalar mesons, i.e. $\pi$, $K$ and $\eta$,
are treated as point-like Goldstone boson. This method has been successfully applied to
the strong decays of heavy-light mesons, charmed
and strange baryons~\cite{Zhong:2008kd,Zhong:2010vq,Zhong:2009sk,
Zhong:2007gp,Liu:2012sj,Xiao:2013xi,Nagahiro:2016nsx,Wang:2017hej,Xiao:2014ura,Xiao:2017udy,Yao:2018jmc}.
The nonrelativistic form of the effective quark-pseudoscalar-meson interactions might be
described by~\cite{Zhong:2008kd,Zhong:2010vq,Zhong:2009sk,
Zhong:2007gp,Liu:2012sj,Xiao:2013xi,Wang:2017hej,Xiao:2014ura}:
\begin{eqnarray}\label{ccpk}
H_{m}^{nr}=\sum_j\left[ \mathcal{G} \vsig_j \cdot \textbf{q}
+h \vsig_j\cdot \textbf{p}_j\right]I_j
e^{-i\mathbf{q}\cdot \mathbf{r}_j},
\end{eqnarray}
with $\mathcal{G}\equiv -\left(1+\frac{\omega_m}{E_f+M_f}\right)$, $h\equiv \frac{\omega_m}{2\mu_q}$.
In the above equation, $\vsig_j$ and $\textbf{p}_j$ stand for the
Pauli spin vector and internal momentum operator for the $j$th quark of the initial hadron;
$\omega_m$ and $\mathbf{q}$ stand for the energy and three momenta of the emitted light meson,
respectively; $E_f$ and $M_f$ are the energy and mass of the final heavy baryon;
$\mu_q$ is a reduced mass given by $1/\mu_q=1/m_j+1/m'_j$ with
$m_j$ and $m'_j$ for the masses of the $j$th quark in the initial and
final hadrons, respectively; and $I_j$ is the isospin operator associated
with the pseudoscalar mesons, which have been defined in Refs.~\cite{Zhong:2007gp,Zhong:2008kd}.
It should be mentioned that the nonrelativistic form
of quark-pseudoscalar-meson interactions expressed in Eq.~(\ref{ccpk}) is similar to that in Refs.~\cite{Godfrey:1985xj,Koniuk:1979vy,Capstick:2000qj},
except that the factors $\mathcal{G}$ and $h$ in this work have explicit
dependence on the energies of final hadrons.

For a light pseudoscalar meson emission in a strong decay process,
the partial decay width can be calculated with~\cite{Zhong:2008kd, Zhong:2007gp}
\begin{equation}\label{dww}
\Gamma_m=\left(\frac{\delta}{f_m}\right)^2\frac{(E_f+M_f)|\mathbf{q}|}{4\pi
M_i(2J_i+1)} \sum_{J_{fz},J_{iz}}|\mathcal{M}_{J_{fz},J_{iz}}|^2 ,
\end{equation}
where $\mathcal{M}_{J_{fz},J_{iz}}$ corresponds to
the strong  amplitudes. The quantum numbers $J_{iz}$ and $J_{fz}$ stand for the third components of the total
angular momenta of the initial and final heavy baryons,
respectively. $M_i$ is the mass of the initial heavy baryon.
$\delta$ as a global parameter accounts for the
strength of the quark-meson couplings. It has been determined in our previous study of the strong
decays of the charmed baryons and heavy-light mesons
\cite{Zhong:2007gp,Zhong:2008kd}. Here, we fix its value the same as
that in Refs.~\cite{Zhong:2008kd,Zhong:2007gp}, i.e. $\delta=0.557$.

In the calculation, we adopt the same quark model parameter set as that in Refs.~\cite{Wang:2017kfr,Yao:2018jmc}.
The masses of the well-established hadrons used in the calculations are
taken from the Particle Data Group (PDG)~\cite{Tanabashi:2018oca}, and the masses of the undiscovered initial states adopted from
the predictions in Ref.~\cite{Ebert:2011kk}.

\begin{figure}[ht]
\centering \epsfxsize=9.2 cm \epsfbox{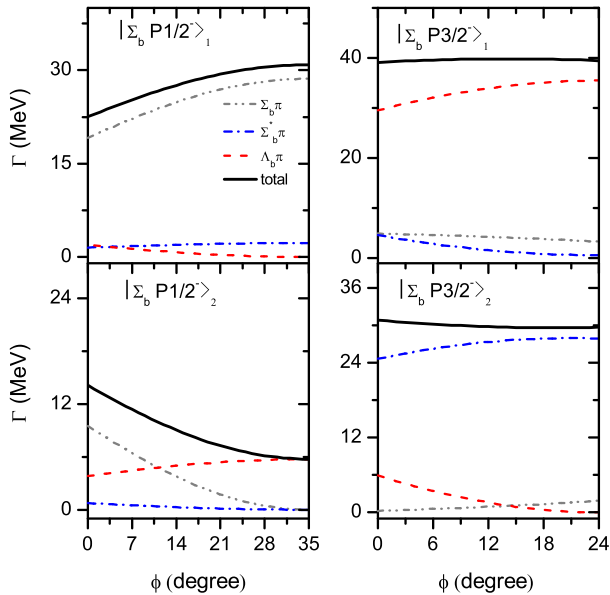}   \vspace{-0.5cm} \caption{Partial and total strong decay widths of the the $J^P = 1/2^-$  and $J^P = 3/2^-$ states with $M = 6097$ MeV in the  $\Sigma_b$
families as a function of mixing angle $\phi$. The solid curves stand for the total widths.  }\label{mixst}
\end{figure}

\begin{table*}[htb]
\begin{center}
\caption{ \label{sigemac} Partial widths (MeV) and branching fractions for the strong decays of the $1P$-wave states in the $j$-$j$ coupling
 scheme compared with that in the $L$-$S$ coupling scheme taken form Ref.~\cite{Wang:2017kfr} in
the $\Sigma_b$ family. The masses of the $P$-wave bottom baryons are adopted from the quark model predictions in Ref.~\cite{Ebert:2011kk}. }
\begin{tabular}{p{1.5cm}|p{2.0cm}p{1.4cm}p{1.4cm}p{1.5cm}|p{2.0cm}p{1.4cm}p{2.4cm}p{1.4cm}p{1.4cm}cccccccccccc}
\hline\hline
              ~~~~ State   &$|J,j\rangle$       & Channel &$\Gamma_{i}$ (MeV)&$\mathcal{B}_i$     &$|^{2S+1}L_{J}\rangle$  &Channel &$\Gamma_{i}$ (MeV)~\cite{Wang:2017kfr}&$\mathcal{B}_i$ ~\cite{Wang:2017kfr}  \\
\hline
$\Sigma_b(6101)$  &$|J=\frac{1}{2}^-,1\rangle$   &$\Lambda_b\pi$ &$\cdot\cdot\cdot$ &$\cdot\cdot\cdot$ &$|^2P_{1/2}  \rangle$&$\Lambda_b\pi$ &1.74  &7.68\%  \\
&                                              &$\Sigma_b\pi$  &28.89    &92.04\%    &                                       &$\Sigma_b\pi$  &19.26 &85.00\% \\
&                                              &$\Sigma_b^*\pi$&2.50      &7.96\%    &            &$\Sigma_b^*\pi$&1.66  &7.33\%\\
&                                              &total          &31.39 &                                 &                &total          &22.66 & \\
\hline

$\Sigma_b(6095)$    &$|J=\frac{1}{2}^-,0\rangle$&$\Lambda_b\pi$ &6.00  &100.0\%                        &$|^4P_{1/2} \rangle$&$\Lambda_b\pi$ &4.00  &28.15\% \\
&                                               &$\Sigma_b\pi$  &$\cdot\cdot\cdot$ &$\cdot\cdot\cdot$& &$\Sigma_b\pi$  &9.50  &66.85\% \\
&                                               &$\Sigma_b^*\pi$&$\cdot\cdot\cdot$ &$\cdot\cdot\cdot$& &$\Sigma_b^*\pi$&0.71  &5.0\%  \\
&                                               &total          &6.00 &              &                &total          &14.21 &\\
\hline

$\Sigma_b(6096)$    &$|J=\frac{3}{2}^-,2\rangle$&$\Lambda_b\pi$ &35.18 &90.07\%                              &$|^2P_{3/2} \rangle$ &$\Lambda_b\pi$ &29.31 &74.60\% \\
&                                               &$\Sigma_b\pi$  &3.25  &8.32\%                          &                                        &$\Sigma_b\pi$  &4.81  &12.24\%        \\
&                                               &$\Sigma_b^*\pi$&0.63  &1.61\%                       &                                        &$\Sigma_b^*\pi$&5.17  &13.16\% \\
&                                               &total          &39.06 &                             &                                         &total          &39.29 &\\
\hline

$\Sigma_b(6087)$ &$|J=\frac{3}{2}^-,1\rangle$&$\Lambda_b\pi$ &$\cdot\cdot\cdot$ &$\cdot\cdot\cdot$ & $|^4P_{3/2}  \rangle$ &$\Lambda_b\pi$ &5.38   &20.46\%\\
&                                            &$\Sigma_b\pi$  &1.47  &5.28\%                         &                &$\Sigma_b\pi$  &0.20   &0.76\%  \\
&                                            &$\Sigma_b^*\pi$&26.39  &94.72\%                          &                &$\Sigma_b^*\pi$&20.71  &78.78\%\\
&                                            &total             &27.86  &                           &                    &total             &26.29  & \\
\hline
$\Sigma_b(6084)$ &$|J=\frac{5}{2}^-,2\rangle$&$\Lambda_b\pi$&31.38 &81.85\% &$|^4P_{5/2} \rangle$ &$\Lambda_b\pi$&31.38 &81.85\% \\
&                                            &$\Sigma_b\pi$ &1.09  &2.84\%                          &                &$\Sigma_b\pi$ &1.09  &2.84\%  \\
&                                            &$\Sigma_b^*\pi$ &5.77&15.05\%                         &                &$\Sigma_b^*\pi$ &5.77&15.05\% \\
&                                            &total           &38.34  &                            &                &total           &38.34  &\\
\hline
\end{tabular}
\end{center}
\end{table*}

\section{Results and discussions}\label{results}

\subsection{$\Sigma_b(6097)^{\pm}$ as the $P$-wave $\Sigma_b$ states}

The measured mass of the $\Sigma_b(6097)$ indicates that it is a good candidate of the $\lambda$-mode $P$-wave excitations.
In the $L$-$S$ coupling scheme, the strong decay properties of the $P$ wave $\Sigma_b$ baryons have
been studied in Ref.~\cite{Wang:2017kfr} (see Table.~\ref{sigemac}). From the decay behaviors, it is found that $\Sigma_b(6097)$
favors the $J^P=3/2^-$ $|\Sigma_b~^2P_{3/2} \rangle$ state or the $J^P=5/2^-$ $|\Sigma_b~^4P_{5/2} \rangle$ state.
Moreover, the possibilities of $\Sigma_b(6097)$ as a candidate of the other $1P$ wave $\Sigma_b$ states
cannot be excluded, because their predicted total widths are close to that of $\Sigma_b(6097)$,
and the $\Lambda_b\pi$ decay mode is allowed.

\subsubsection{$J^P=1/2^-$ states}

In the $j$-$j$ coupling scheme, according to Eq.~(\ref{Rela}) we can obtain two $J^P=1/2^-$ states as mixed states between $|^2P_{1/2} \rangle$ and $|^4P_{1/2} \rangle$:
\begin{eqnarray}
\left|J^P=\frac{1}{2}^-,0\right\rangle &= &-\sqrt{\frac{1}{3}}\left|^2P_{1/2} \right\rangle + \sqrt{\frac{2}{3}}\left|^4P_{1/2} \right\rangle,\\
\left|J^P=\frac{1}{2}^-,1\right\rangle &= &\sqrt{\frac{2}{3}}\left|^2P_{1/2} \right\rangle + \sqrt{\frac{1}{3}}\left|^4P_{1/2}\right \rangle.
\end{eqnarray}
If let $\cos \phi=\sqrt{\frac{2}{3}}$ and $\sin \phi=\sqrt{\frac{1}{3}}$, we get
the mixing angle $\phi\simeq 35^\circ$. The strong decay properties of $|J^P=\frac{1}{2}^-,0\rangle$ and $|J^P=\frac{1}{2}^-,1\rangle$ are calculated within the chiral quark model, our results are presented in Table.~\ref{sigemac}. It is seen that the $|J^P=\frac{1}{2}^-,0\rangle$ is a narrow state with a width of a few MeV, and mainly decays into $\Lambda_b \pi$; while the $|J^P=\frac{1}{2}^-,1\rangle$ state has a width of $\sim 30$ MeV, and dominantly decays into $\Sigma_b\pi$. For a comparison, in Table.~\ref{sigemac} we also list the strong decay properties of the $J^P=1/2^-$ states $|^2P _{1/2} \rangle$ and $|^4P_{1/2}\rangle$ obtained from the $L$-$S$ coupling scheme in Ref.~\cite{Wang:2017kfr}. It is found that
the decay properties of $|J^P=\frac{1}{2}^-,1\rangle$ in the  $j$-$j$ coupling scheme are similar to those of $|^2P_{1/2}\rangle$ in the $L$-$S$ coupling scheme, for the strong decays of $|J^P=\frac{1}{2}^-,1\rangle$ are governed by the component of $|^2P_{1/2}\rangle$. However, the strong decay properties of $|J^P=\frac{1}{2}^-,0\rangle$ are very different from those in the $L$-$S$ coupling scheme.
The decay properties of the $J^P=1/2^-$ states are inconsistent with the observations of $\Sigma_b(6097)$, thus, the $\Sigma_b(6097)$ as the $J^P=1/2^-$ states should be excluded.

It is interesting to note that the $\Omega_c(3000)$ resonance may be explained as a mixed state $|P_\lambda~{\frac{1}{2}}^-\rangle_1=\cos \phi|^2P_{1/2} \rangle+\sin \phi|^4P_{1/2} \rangle$
with the mixing angle $\phi\simeq 24^\circ$~\cite{Wang:2017hej}, which is close to
$\phi\simeq 35^\circ$ determined by the heavy-quark symmetry in the $j$-$j$ coupling scheme,
but not actually equal to that determined by the heavy-quark symmetry.
This indicates that on the one hand the heavy-quark symmetry requires the physical states might
be mixed states between the states with the same $J^P$ in the $L$-$S$ coupling scheme,
on the other hand because the heavy-quark symmetry is only an approximation, the physical mixing angle should be slightly different
from that determined within the $j$-$j$ coupling scheme.
Thus, the physical $P$-wave singly-heavy baryon states might
be mixed states between $|^2P _{J} \rangle$ and $|^4P_J\rangle$ in the $L$-$S$ coupling scheme, i.e.,
\begin{equation}\label{mixd}
\left(\begin{array}{c}| P~{J^P}\rangle_1\cr |  P~{J^P}\rangle_2
\end{array}\right)=\left(\begin{array}{cc} \cos\phi & \sin\phi \cr -\sin\phi &\cos\phi
\end{array}\right)
\left(\begin{array}{c} |^2P_{J}
\rangle \cr |^4P_{J}\rangle
\end{array}\right),
\end{equation}
and the mixing angle $\phi$ may range from the value of the $L$-$S$ coupling scheme ($\phi=0^\circ$) to
that of the $j$-$j$ coupling scheme.
Specially, for the $P$-wave states with $J^P=1/2^-$ the mixing angle $\phi\in (0,35^\circ)$,
while for the $P$-wave states with $J^P=3/2^-$ the mixing angle $\phi\in (0,24^\circ)$.

According to the mixing scheme defined in Eq.~(\ref{mixd}),
the strong decay widths of the $J^P=1/2^-$ states in the $\Sigma_b$ family as a function of the mixing angle $\phi$ are shown
in Fig.~\ref{mixst}. It is found that the mixed state
$|\Sigma_b~P{\frac{1}{2}^-}\rangle_1$ has a width of $\sim 20-30$ MeV, its decays are dominated by the $\Sigma_b \pi$ channel,
the decay properties are less sensitive to the mixing angle. If the mixing angle is less than
the angle $\phi\simeq 35^\circ$ obtained in the heavy quark limit, there is a small decay rate into the $\Lambda_b\pi$ channel.
The other mixed state $|\Sigma_b~P{\frac{1}{2}^-}\rangle_2$ has a relatively narrower width, which is sensitive to the mixing angle.
If we take the same mixing angle $\phi\simeq 24^\circ$ as that of $\Omega_c(3000)$~\cite{Wang:2017hej},
except for the main decay mode $\Lambda_b\pi$, the $\Sigma_b\pi$ may play an obvious role in its strong decays.
The decay properties of $|\Sigma_b~P{\frac{1}{2}^-}\rangle_1$ and $|\Sigma_b~P{\frac{1}{2}^-}\rangle_2$
are inconsistent with the observations of $\Sigma_b(6097)$.

\subsubsection{$J^P=3/2^-$ states}

For the $J^P=3/2^-$ states $|J^P=\frac{3}{2}^-,1\rangle$ and $|J^P=\frac{3}{2}^-,2\rangle$ in the $j$-$j$ coupling scheme, according to Eq.~(\ref{Rela}) they can relate to the states in the $L$-$S$ coupling scheme:
\begin{eqnarray}\label{a}
\left|J^P=\frac{3}{2}^-,1\right\rangle &= &-\sqrt{\frac{1}{6}}\left|^2P_{3/2} \right\rangle + \sqrt{\frac{5}{6}}\left|^4P_{3/2} \right\rangle,
\end{eqnarray}
\begin{eqnarray}\label{b}
\left|J^P=\frac{3}{2}^-,2\right\rangle &= &\sqrt{\frac{5}{6}}\left|^2P_{3/2} \right\rangle + \sqrt{\frac{1}{6}}\left|^4P_{3/2}\right \rangle.
\end{eqnarray}
If let $\cos \phi=\sqrt{\frac{5}{6}}$ and $\sin \phi=\sqrt{\frac{1}{6}}$, we get
the mixing angle $\phi\simeq 24^\circ$. The decay properties of $|J^P=\frac{3}{2}^-,1\rangle$ and $|J^P=\frac{3}{2}^-,2\rangle$
in the $\Sigma_b$ family are calculated within the chiral quark model. Our results are listed in  Table.~\ref{sigemac} as well.
From the Table.~\ref{sigemac}, it is found that the $|J^P=\frac{3}{2}^-,1\rangle$ state has a width of $\sim 28$ MeV,
and dominantly decays into the $\Sigma_b^*\pi$ channel; while the $|J^P=\frac{3}{2}^-,2\rangle$ state has
a relatively broad width of $\sim 39$ MeV, and dominantly decays into the $\Lambda_b\pi$ channel.
Both the decay mode and width
indicate that $\Sigma_b(6097)$ is a good candidate of the light spin $j=2$ state $|J^P=\frac{3}{2}^-,2\rangle$.
The partial width ratio between $\Lambda_b \pi$ and $\Sigma_b\pi$ for $|J^P=\frac{3}{2}^-,2\rangle$
is predicted to be
\begin{eqnarray}\label{b}
\frac{\Gamma[\Lambda_b\pi]}{\Gamma[\Sigma_b\pi]}\simeq 12,
\end{eqnarray}
which can be tested in future experiments. It should be mentioned that the main decay properties, such as the total decay width and dominant decay modes, of the $J^P=3/2^-$ states $|J^P=\frac{3}{2}^-,1\rangle$ and $|J^P=\frac{3}{2}^-,2\rangle$ in the $j$-$j$ coupling scheme are similar to those of $|^4P _{3/2} \rangle$ and $|^2P_{3/2}\rangle$ in the $L$-$S$ coupling scheme, respectively (see Table.~\ref{Xipcb}).

Considering the mixing angle of the physical states with $J^P=3/2^-$ may have some deviations from the $j$-$j$ couplings, adopting the mixing scheme defined in Eq.~(\ref{mixd}) we also plot the strong decay width as a function of mixing angle $\phi$ in Fig.~\ref{mixst} for a reference. It is shown that when $\phi$ varies in $0\sim 24^\circ$, the total widths of both $|\Sigma_b~P ~\frac{3}{2}^-\rangle_1$ and $|\Sigma_b~P ~\frac{3}{2}^-\rangle_2$ are less sensitive to the mixing angle, however, the partial width ratio of $\Gamma[\Lambda_b\pi]/\Gamma[\Sigma_b^*\pi]$ is very sensitive to the mixing angle. The measurements of the ratio $\Gamma[\Lambda_b\pi]/\Gamma[\Sigma_b^*\pi]$ might be helpful to determine the mixing angle.

\subsubsection{$J^P=5/2^-$ state}

For the $J^P=5/2^-$ state, the $L$-$S$ coupling scheme is equal to the $j$-$j$ coupling scheme, i.e.,
\begin{eqnarray}
\left|J^P=\frac{5}{2}^-,2\right\rangle &= &\left|^4P_{5/2} \right\rangle.
\end{eqnarray}
From Table~\ref{sigemac}, it is seen that both the total decay width and the dominant
$\Lambda_b \pi$ decay mode of $|J^P=\frac{5}{2}^-,2\rangle$ also favor the observations of $\Sigma_b(6097)$.
The decay rate of the $J^P=5/2^-$ state into $\Sigma_b^*\pi$ is sizable as well.
The partial width ratio between the two main decay channels $\Lambda_b \pi$ and $\Sigma_b^*\pi$
is predicted to be
\begin{eqnarray}\label{qq}
\frac{\Gamma[\Lambda_b\pi]}{\Gamma[\Sigma_b^*\pi]}\simeq 5.
\end{eqnarray}
If the $\Sigma_b(6097)$ resonance is the $J^P=5/2^-$ state indeed, it should be observed in the $\Sigma_b^*\pi$ channel as well.

As a whole, $\Sigma_b(6097)$ seems to favor the $j=2$ states with $J^P=3/2^-$ or $J^P=5/2^-$. To distinguish
these two states, more observations in the $\Sigma_b\pi$ and $\Sigma_b^*\pi$ channels are needed in future experiments.
If $\Sigma_b(6097)$ corresponds to $|J^P=\frac{3}{2}^-,2\rangle$, it should be observed in the $\Sigma_b\pi$ channel.
On the other hand, if $\Sigma_b(6097)$ is the $J^P=5/2^-$ state, it should be observed in the $\Sigma_b^*\pi$ channel.

\begin{figure}[ht]
\centering \epsfxsize=9.2 cm \epsfbox{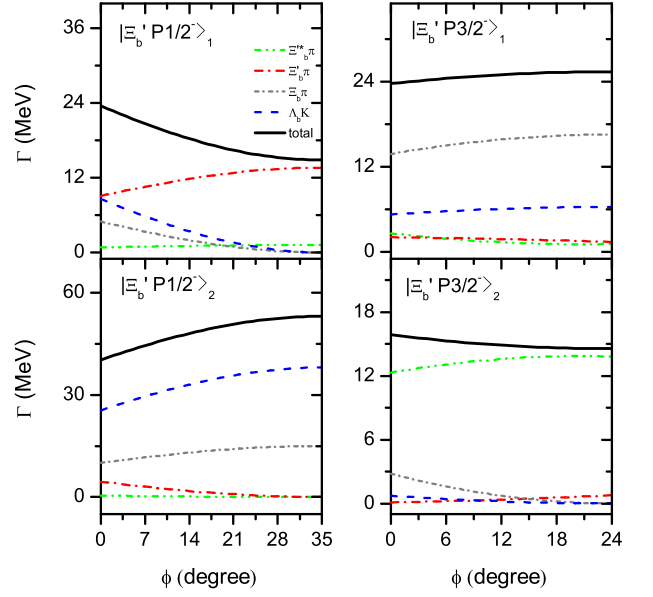}   \vspace{-0.5cm} \caption{Partial and total strong decay widths of  the $J^P = 1/2^-$  and $J^P = 3/2^-$ states with $M = 6227$ MeV in the $\Xi'_b$
families as  a function of mixing angle $\phi$. The solid curves stand for the total widths.}\label{KSBH}
\end{figure}

\begin{table*}[htb]
\begin{center}
\caption{ \label{Xipcb} Partial widths (MeV) and branching fractions for the strong decays of the $1P$-wave states in the $j$-$j$ coupling scheme compared with that in the $L$-$S$ coupling scheme taken form Ref.~\cite{Wang:2017kfr} in
the $\Xi_b'$ family. The masses of the $P$-wave bottom baryons are adopted from the quark model predictions in Ref.~\cite{Ebert:2011kk}.}
\begin{tabular}{p{1.5cm}|p{2.0cm}p{1.4cm}p{1.4cm}p{1.4cm}|p{2.0cm}p{1.8cm}p{2.4cm}p{1.4cm}p{1.4cm}cccccccccccc}
\hline\hline
~~~~State   & $|J,j\rangle$       &Channel         &$\Gamma_{i}$ (MeV)    &$\mathcal{B}_i$  &$|^{2S+1}L_J\rangle$       &Channel         &$\Gamma_{i}$ (MeV)~\cite{Wang:2017kfr}    &$\mathcal{B}_i$~\cite{Wang:2017kfr} \\
\hline

$\Xi'_b(6233)$   &$|J=\frac{1}{2}^-,1\rangle$ &$\Lambda_bK$  &$\cdot\cdot\cdot$    &$\cdot\cdot\cdot$ &$|^2P_{1/2} \rangle$  &$\Lambda_bK$        &12.11        &44.77\%   \\
                            &                 &$\Xi_b\pi$    &$\cdot\cdot\cdot$    &$\cdot\cdot\cdot$                        &                 &$\Xi_b\pi$          &4.77        &17.63\%   \\
                              &   &$\Xi^\prime_b\pi$      &13.85    &90.76\% &           &$\Xi^\prime_b\pi$         &9.23       &34.12\% \\
                               &  &$\Xi^\prime_b(5945)\pi$   &1.41 &9.24\% &           &$\Xi^\prime_b(5945)\pi$   &0.94        &3.48\%\\
                               &          &total            &15.26       &                  &           &total                     &27.05       &            \\
\hline

$\Xi'_b(6227)$     &$|J=\frac{1}{2}^-,0\rangle$   &$\Lambda_bK$   &38.07  &71.71\%   &$|^4P_{1/2}  \rangle$ &$\Lambda_bK$    &17.28   &53.60\%  \\
                           &                 &$\Xi_b\pi$      &15.02     &28.29\%    &                                       &$\Xi_b\pi$          &10.01        &31.05\%         \\
                          &           &$\Xi^\prime_b\pi$         &$\cdot\cdot\cdot$       &$\cdot\cdot\cdot$   &     &$\Xi^\prime_b\pi$         &4.54       &14.08\%\\
                               &           &$\Xi^\prime_b(5945)\pi$   &$\cdot\cdot\cdot$  &$\cdot\cdot\cdot$   &      &$\Xi^\prime_b(5945)\pi$   &0.41        &1.27\%       \\
                               &                   &total              &53.09       &                       &                          &total              &32.24       &  \\
\hline

$\Xi'_b(6234)$         &$|J=\frac{3}{2}^-,2\rangle$   &$\Lambda_bK$        &7.31   &26.02\%   &$|^2P_{3/2}  \rangle$   &$\Lambda_bK$    &4.14    &17.14\%  \\
                          &                 &$\Xi_b\pi$          &17.89        &63.69\%         &                 &$\Xi_b\pi$          &14.91        &61.74\%             \\
                         &           &$\Xi^\prime_b\pi$         &1.60       &5.70\%         &           &$\Xi^\prime_b\pi$         &2.37       &9.81\%        \\
                           &           &$\Xi^\prime_b(5945)\pi$   &1.29        &4.59\%          &           &$\Xi^\prime_b(5945)\pi$   &2.73        &11.30\%       \\
                               &                  &total             &28.09       & &                  &total              &24.15       & \\

\hline
$\Xi'_b(6224)$     &$|J=\frac{3}{2}^-,1\rangle$   &$\Lambda_bK$  &$\cdot\cdot\cdot$   &$\cdot\cdot\cdot$ & $|^4P_{3/2}  \rangle$    &$\Lambda_bK$     &0.98        &6.19\%         \\
                          &                 &$\Xi_b\pi$          &$\cdot\cdot\cdot$   &$\cdot\cdot\cdot$                           &                 &$\Xi_b\pi$       &2.67        &16.87\% \\
                           &           &$\Xi^\prime_b\pi$         &0.72       &5.02\%                                        &           &$\Xi^\prime_b\pi$         &0.10       &0.63\%\\
                            &           &$\Xi^\prime_b(5945)\pi$   &13.61       &94.98\%                                      &           &$\Xi^\prime_b(5945)\pi$   &12.08        &76.31\%      \\
                           &                   &total              &14.33       &                                           &                   &total              &15.83       &  \\
\hline
$\Xi'_b(6226)$         &$|J=\frac{5}{2}^-,2\rangle$   &$\Lambda_bK$        &4.20        &17.22\%     &$|^4P_{5/2} \rangle$  &$\Lambda_bK$        &4.20        &17.22\%         \\
                             &                 &$\Xi_b\pi$          &16.37        &67.12\%                       &                 &$\Xi_b\pi$          &16.37        &67.12\%\\
                          &           &$\Xi^\prime_b\pi$         &0.60       &2.46\%                              &           &$\Xi^\prime_b\pi$         &0.60       &2.46\%\\\
                            &           &$\Xi^\prime_b(5945)\pi$   &3.22        &13.20\%                          &           &$\Xi^\prime_b(5945)\pi$   &3.22        &13.20\%\\
                             &                   &total             &24.39       &                              &                   &total             &24.39       &  \\
\hline\hline
\end{tabular}
\end{center}
\end{table*}

\subsection{ $\Xi'_b$(6227)$^-$ as the $P$-wave $\Xi_b'$ states}

The measured mass of $\Xi_b(6227)$ indicates that it is a good candidate of the $\lambda$-mode
$P$-wave excitations~\cite{Ebert:2011kk}.
Within the $L$-$S$ coupling scheme, the strong decay behaviors of the $P$ wave $\Xi'_b$ states have
been studied in Ref.~\cite{Wang:2017kfr}. It is found that $\Xi_b(6227)$ favors the $J^P=3/2^-$ $|\Xi'_b~^2P_{3/2} \rangle$ state or $J^P=5/2^-$ $|\Xi'_b~^4P_{5/2} \rangle$ state. Moreover, due to the theoretical uncertainties and lack of the branching ratio information, the possibilities of the $\Xi_b(6227)$ as other $P$ wave $\Xi'_b$ states cannot be excluded.

In this work, we study the strong decays of the $\lambda$-mode $P$-wave $\Xi'_b$ state in the $j$-$j$ coupling scheme, and our results are listed in Table~\ref{Xipcb}.

\subsubsection{$J^P=1/2^-$ states}

It is found that the $J^P=1/2^-$ state $|J^P=\frac{1}{2}^-,0\rangle$ has a width of 51 MeV, this state mainly decays into $\Lambda_bK$ and $\Xi_b\pi$ channels. The decay width of $|J^P=\frac{1}{2}^-,0\rangle$ is about a factor 3 larger than that of $\Xi_b(6227)$. While the other $J^P=1/2^-$ state $|J^P=\frac{1}{2}^-,1\rangle$ has a width of 15 MeV, and dominantly decays into $\Xi_b'\pi$ channel.
Although the decay width of $|J^P=\frac{1}{2}^-,1\rangle$ is close to that of $\Xi_b(6227)$, the decay modes are inconsistent with the observations.
For a comparison, in Table.~\ref{Xipcb} we also list the strong decay properties of the $J^P=1/2^-$ states $|^2P _{1/2} \rangle$ and $|^4P_{1/2}\rangle$ which are calculated within the $L$-$S$ coupling scheme in Ref.~\cite{Wang:2017kfr}. The predictions show obvious differences between the $j$-$j$ and $L$-$S$ coupling schemes (see Table.~\ref{Xipcb}).

Considering the mixing angle between $|^2P _{1/2} \rangle$ and $|^4P_{1/2}\rangle$ for the physical states with $J^P=1/2^-$ may have some deviations from the $j$-$j$ coupling scheme, the strong decay widths $|\Xi_b'~P{\frac{1}{2}^-}\rangle_1$ and $|\Xi_b'~P{\frac{1}{2}^-}\rangle_2$ as a function of mixing angle $\phi$ are shown in Fig.~\ref{KSBH}. It is found that the decay properties of $|\Xi_b'~P{\frac{1}{2}^-}\rangle_2$ are less sensitive to the mixing angle $\phi$. However, the decay properties of $|\Xi_b'~P{\frac{1}{2}^-}\rangle_1$ shows some sensitivities to the mixing angle.
For example, if one adopts the mixing angle $\phi\simeq 35^\circ$ in the heavy quark limit the $\Lambda_bK$ mode for $|\Xi_b'~P{\frac{1}{2}^-}\rangle_1$ is forbidden, while if $\phi\simeq 15^\circ$ the decay rate into $\Lambda_bK$ is sizable.
From the decay properties shown in Fig.~\ref{KSBH}, one can find that the $\Xi'_b(6227)$ do not favor the mixed states $|\Xi'_b~P {\frac{1}{2}^-}\rangle_1$ and $|\Xi'_b~P {\frac{1}{2}^-}\rangle_2$.

\subsubsection{$J^P=3/2^-$ states}

For the $J^P=3/2^-$ state $|J^P=\frac{3}{2}^-,1\rangle$ in the $\Xi_b'$ family, it has a width of $\sim$14 MeV,
and dominantly decays into $\Xi_b'(5945) \pi$, the decay mode is inconsistent with the observations of $\Xi'_b(6227)$.
While the other $J^P=3/2^-$ state $|J^P=\frac{3}{2}^-,2\rangle$ has a width of $\sim$24 MeV, and mainly decays into
$\Lambda_b K$ and $\Xi_b \pi$ final states, and the partial width ratio between these two channels is predicted to be
\begin{eqnarray}\label{qq}
\frac{\Gamma[\Lambda_b K]}{\Gamma[\Xi_b \pi]}\simeq 0.36.
\end{eqnarray}
This ratio predicted by us is obviously smaller than the $^3P_0$ model prediction $\frac{\Gamma[\Lambda_b K]}{\Gamma[\Xi_b \pi]}\simeq 0.89$ in Ref.~\cite{Chen:2018orb}.
Both the decay modes and width of $|J^P=\frac{3}{2}^-,2\rangle$ are consistent with the observations of $\Xi_b(6227)$. It should be mentioned that the strong decay properties of the $J^P=3/2^-$ $\Xi_b'$ states $|J^P=\frac{3}{2}^-,1\rangle$ and $|J^P=\frac{3}{2}^-,2\rangle$ in the $j$-$j$ coupling scheme are similar to those of $|^4P _{3/2} \rangle$ and $|^2P_{3/2}\rangle$ in the $L$-$S$ coupling scheme, respectively (see Table.~\ref{Xipcb}).

The mixing angle of the physical states with $J^P=3/2^-$ may lie between the $L$-$S$ and $j$-$j$ coupling limit,
we show the strong decay widths as a function of mixing angle $\phi$ from $0^\circ\sim 24^\circ$ in Fig.~\ref{KSBH}.  It can be found that the deviation from the $j$-$j$ coupling mixing has small influence on the strong decay behaviors, and our conclusions remain.
It should mentioned that the recent QCD sum rule analysis also suggested that $\Xi_b(6227)$ may be a $J^P=3/2^-$ state~\cite{Agaev:2017nn}.

\subsubsection{$J^P=5/2^-$ state}

For the $|J^P=\frac{5}{2}^-,2\rangle$ state, the $L$-$S$ coupling scheme is equal to the $j$-$j$ coupling scheme. The results indicate that the $\Xi'_b(6227)$ can be assigned as the $|J^P=\frac{5}{2}^-,2\rangle$ state as well.
The partial width ratio between $\Lambda_b K$ and $\Xi_b \pi$ is predicted to be
\begin{eqnarray}\label{qq}
\frac{\Gamma[\Lambda_b K]}{\Gamma[\Xi_b \pi]}\simeq 0.25,
\end{eqnarray}
which is slightly smaller than that for $|J^P=\frac{3}{2}^-,2\rangle$. The ratio predicted by us is obviously smaller than the $^3P_0$ model prediction $\frac{\Gamma[\Lambda_b K]}{\Gamma[\Xi_b \pi]}\simeq 0.94$ in Ref.~\cite{Chen:2018orb}.

As a whole, $\Xi'_b(6227)$ seems to favor the $j=2$ states with $J^P=3/2^-$ or $J^P=5/2^-$.
Our conclusion is also consistent with the calculations of the $^3P_0$ model and QCD sum rule~\cite{Aliev:2018lcs,Chen:2018orb}.
To further understand the nature of $\Xi'_b(6227)$, more measurements, such as the ratio
$\Gamma[\Lambda_b K]/\Gamma[\Xi_b \pi]$, are suggested to be carried out in future experiments.

\begin{figure}[ht]
\centering \epsfxsize=9.2 cm \epsfbox{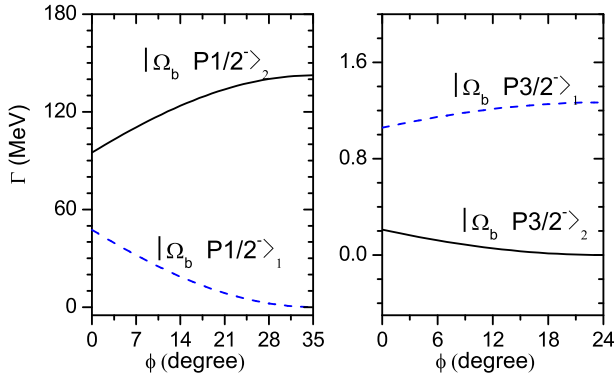}   \vspace{-0.5cm} \caption{Partial decay widths of the $\Xi_b K$ channel for the $J^P = 1/2^-$ and $J^P = 3/2^-$ $\Omega_b$ states as a function of mixing angle $\phi$. The resonance mass is taken the quark model prediction $M = 6330$ MeV.}\label{omigebH}
\end{figure}

\subsection{The $P$-wave $\Omega_b$ states}

\begin{table*}[htp]
\begin{center}
\caption{\label{owmigeP}   Partial widths (MeV) and branching fractions for the strong decays of the $1P$-wave states in the $j$-$j$ coupling scheme compared with that in the $L$-$S$ coupling scheme taken form Ref.~\cite{Wang:2017kfr} in
the $\Omega_b$ family. The masses of the $P$-wave bottom baryons are adopted from the quark model predictions in Ref.~\cite{Ebert:2011kk}.}
\begin{tabular}{p{2.5cm}|p{2.5cm}p{2.5cm}|p{2.5cm}p{3.5cm}p{2.5cm}p{2.0cm}p{1.5cm}ccccccccccccccccccccccccccc}\hline\hline
State ~\cite{Ebert:2011kk}            &$~~~|J^P,j\rangle$   &~~$\Gamma[\Xi_bK]$ (MeV) ~~    &$|^{2S+1}L_J\rangle$       &~~$\Gamma[\Xi_bK]$ (MeV)~\cite{Wang:2017kfr} \\ \hline
$\Omega_b(6339)$      &$|J=\frac{1}{2}^-,1\rangle$   &~~~~~~$\cdot\cdot\cdot$      &$|^2P_{1/2} \rangle$     &~~~~~~49           \\
$\Omega_b(6330)$      &$|J=\frac{1}{2}^-,0\rangle$   &~~~~~~143                  &$|^4P_{1/2}\rangle$     &~~~~~~95        \\
$\Omega_b(6340)$      &$|J=\frac{3}{2}^-,2\rangle$   &~~~~~~2.2                    &$|^2P_{3/2} \rangle$     &~~~~~~1.8     \\
$\Omega_b(6331)$     &$|J=\frac{3}{2}^-,1\rangle$   &~~~~~~$\cdot\cdot\cdot$       &$|^4P_{3/2} \rangle$     &~~~~~~0.22\\
$\Omega_b(6334)$      &$|J=\frac{5}{2}^-,2\rangle$   &~~~~~~1.6                    &$|^4P_{5/2} \rangle$     &~~~~~~1.6\\
\hline
\end{tabular}
\end{center}
\end{table*}

For the $1P$-wave $\Omega_b$ states, there are no signals from experiments. By taking the predicted masses from RQM~\cite{Ebert:2011kk}, their strong decays in $j$-$j$ coupling scheme are calculated and shown in Table~\ref{owmigeP}. For the $J^P=1/2^-$ state $|J^P=\frac{1}{2}^-,0\rangle$, a large decay width of 148 MeV is obtained, its decays are governed by the $\Xi_b K$ mode. The decay mode $\Xi_b K$ for the $j=1$ states $|J^P=\frac{1}{2}^-,1\rangle$ and $|J^P=\frac{3}{2}^-,1\rangle$ are forbidden, thus, these states should be very narrow states. The two $j=2$ states $|J^P=\frac{3}{2}^-,2\rangle$ and $|J^P=\frac{5}{2}^-,2\rangle$ are predicted to be very narrow states with a width of $\sim1-2$ MeV, which might be found in future experiments. The strong decay properties of the $1P$-wave $\Omega_b$ states were also studied in Ref.~\cite{Chen:2018vuc,Agaev:2017nn}. Our main results are consistent with the predictions in Ref.~\cite{Chen:2018vuc}. For a comparison, the results calculated from the $L$-$S$ coupling scheme~\cite{Wang:2017kfr} are also listed in Table.~\ref{owmigeP}. It is found that the strong decay properties of $|J=\frac{1}{2}^-,0\rangle$ and $|J=\frac{3}{2}^-,2\rangle$ are similar to those of $|^4P_{1/2}\rangle$ and $|^2P_{3/2} \rangle$ in the $L$-$S$ coupling scheme, respectively. However, the decay properties of $|J=\frac{1}{2}^-,1\rangle$ and $|J=\frac{3}{2}^-,1\rangle$ are very different from those in the $L$-$S$ coupling scheme.

The physical states with $J^P=1/2^-$ as mixed states between $|^2P _{1/2} \rangle$ and $|^4P_{1/2}\rangle$,
the mixing angle may not be the ideal angle $\phi=35^\circ$ predicted in the heavy quark limit.
Taking the mixing scheme as defined in Eq.~(\ref{mixd}), the strong decay widths of the $J^P=1/2^-$ states as a function of the mixing angle $\phi$  are presented in Fig.~\ref{omigebH}. When the mixing angle lies in $0^\circ\sim 35^\circ$, the width of $|\Omega_b~P {\frac{1}{2}^-}\rangle_1$ is relatively small, while $|\Omega_b~P {\frac{1}{2}^-}\rangle_2$ is a broad state with a width of $\sim 120\pm 20$ MeV. For the $J^P=3/2^-$ states, when the mixing angle varies from $0^\circ$ to $24^\circ$, the $|\Omega_b~P {\frac{3}{2}^-}\rangle_1$ state is a narrow state with a width of $\sim 1$ MeV, while the partial width of $|\Omega_b~P {\frac{3}{2}^-}\rangle_2$ into $\Xi_b K$ channel may be less than 0.2 MeV.
The $J^P=3/2^-$ states might be found in the $\Xi_b K$ channel in future experiments.

\section{Summary}\label{suma}

In the $j$-$j$ coupling scheme, the strong  decays of the low-lying
$P$-wave singly bottom heavy baryons belonging to $\mathbf{6}_F$ are studied within the chiral quark model.
Our results show that the newly observed resonances $\Sigma_b(6097)^{\pm}$ can be assigned
as the light spin $j=2$ states with spin-parity numbers $J^P=3/2^-$ or $J^P=5/2^-$. If $\Sigma_b(6097)$ corresponds to $|J^P=\frac{3}{2}^-,2\rangle$, it might be observed in the $\Sigma_b\pi$ channel, while if $\Sigma_b(6097)$ is the $J^P=5/2^-$ state, it might be observed in the $\Sigma_b^*\pi$ channel.

The newly observed resonance $\Xi_b(6227)^-$ favors the light spin $j=2$ states with spin-parity numbers $J^P=3/2^-$ or $J^P=5/2^-$ in the $\Xi_b'$ family. $\Xi_b(6227)^-$ may be the strange partner of $\Sigma_b(6097)^{\pm}$. The $J^P=3/2^-$ state $|J^P=\frac{3}{2}^-,2\rangle$ and the $J^P=5/2^-$ state $|J^P=\frac{5}{2}^-,2\rangle$ in the $\Xi_b'$ family have similar strong decay properties. In order to identify them, angular distributions of their decays in either strong decay modes or radiative transitions should be needed~\cite{Wang:2017kfr}.

Considering the heavy quark symmetry, the $J^P=1/2^-$ $P$-wave $\Omega_b$ state $|J^P=\frac{1}{2}^-,0\rangle$ might be a broad state with a width of
$\sim 150$ MeV. The two $j=2$ states $|J^P=\frac{3}{2}^-,2\rangle$ and $|J^P=\frac{5}{2}^-,2\rangle$ are predicted to be very narrow states with a width of $\sim1-2$ MeV, which might be found in future experiments. The $\Xi_b K$ decay mode for the $j=1$ states $|J^P=\frac{1}{2}^-,1\rangle$ and $|J^P=\frac{3}{2}^-,1\rangle$ are forbidden in the heavy quark limit, thus these two states should be very narrow states.
To looking for the missing $P$-wave $\Omega_b$ states, the $\Xi_b K$ decay mode is worth to observing.

\section*{  Acknowledgments }

This work is supported, in part, by the National Natural Science Foundation of China under Grants No.11775078,
No.U1832173, and No. 11705056.

\bibliographystyle{unsrt}

\end{document}